\documentclass[preprint,english,aps,prb,showpacs]{revtex4}
\usepackage{graphicx}
\usepackage{babel}
\usepackage{amsfonts}

\def\pref#1{(\ref{#1})}
\def\prefs#1#2{(\ref{#1}-\ref{#2})}

\begin{document}

\title{Parafermionic states in rotating Bose-Einstein condensates}

\author{N.~Regnault}

\affiliation{Laboratoire Pierre Aigrain, D\'epartement de
Physique, 24, rue Lhomond, 75005 Paris,
 France}

\author{Th.~Jolicoeur}

\affiliation{Laboratoire Pierre Aigrain, D\'epartement de
Physique, 24, rue Lhomond, 75005 Paris,
 France}
\affiliation{Laboratoire de Physique Th\'eorique et Mod\`eles Statistiques, Universit\'e Paris-Sud, 91405 Orsay, France.}


\begin{abstract}
We investigate possible parafermionic states in rapidly rotating ultracold bosonic atomic gases at lowest Landau level filling factor $\nu=k/2$. We study how the system size and interactions act upon the overlap between the true ground state and a candidate Read-Rezayi state. We also consider the quasihole states which are expected to display non-Abelian statistics. We numerically evaluate the degeneracy of these states and show agreement with a formula given by E. Ardonne. We compute the overlaps between low-lying exact eigenstates and quasihole candidate wavefunctions. We discuss the validity of the parafermion description as a function of the filling factor.
\end{abstract}


\maketitle

\section{Introduction}

Rotating Bose-Einstein condensates display a wealth of interesting
 physics. One of the most striking achievements in this field is the
observation of the Abrikosov lattice of 
vortices~\cite{Madison00,Aboshaeer01}. With increasing rotation speed, it
has been predicted that this lattice will melt and is replaced by
more exotic quantum phases. When the rotation
frequency is close to the harmonic trapping frequency and strong
confinement is applied along the rotation axis,
 strongly correlated states belonging to the family of quantum
Hall liquid states should
appear~\cite{Wilkin98,Wilkin00,Cooper01,Regnault03}.

Here we study bosonic atoms with only one hyperfine
species (i.e. spinless bosons) in such regime. We assume that the
temperature is low and the interactions are weak enough so that
the lowest Landau level (LLL) approximation is valid. The system
may then display the fractional quantum
Hall effect (FQHE) as in two dimensional electron systems (2DES)
under strong magnetic field. The Coulomb interaction is
replaced by the $s$-wave scattering
between the ultracold atoms. An analog of the filling factor $\nu$ for 2DES
can be defined~: indeed $\nu=N/N_\phi$ is the ratio between the number of
atoms $N$ and the number of vortices $N_\phi$ that would be
present in the system if it was a Bose condensate. The quantity $N_\phi$ is the
equivalent of the number of flux quanta in 2DES systems. In this
regime, several fractions have been predicted. The most prominent
one~\cite{Wilkin98} appears at $\nu=1/2$, for which the Laughlin
state is the exact ground state~\cite{Laughlin83}. Evidences for
other fractions from the Jain principal sequence $\nu=p/(p+1)$ such as
$\nu=2/3$ and $\nu=3/4$ have been pointed out~\cite{Regnault03,Chiachen05} 
and can be understood within the composite fermion
theory\cite{Jain89}.

Hierarchical quantum Hall states are not the only interesting states that have
been predicted below the critical filling factor where the lattice
of vortices melt~\cite{Cooper01}. Due to the bosonic
statistics, if we assume that the equivalent cyclotron gap is
large enough, we can have filling factors greater than one and
still stay entirely in the LLL. Within this  hypothesis, even more exotic
states  should appear for fractions $\nu=k/2$. The first one is
the Moore-Read (MR) state~\cite{Moore91} (or Pfaffian state) that
should occur~\cite{Wilkin00} at $\nu=1$ ($k=2$). This state was
first introduced to explain the fermionic fraction $\nu=5/2$ in 2DES. Higher
$k$ values are associated to the so-called Read-Rezayi (RR)
states~\cite{Read99, Cooper01}. Because of the parafermionic
behavior of these states, their excitations have surprising
non-Abelian statistics. So far, there are no well-established physical situation
where these states play a role. The original suggestion by Read and Rezayi is that
they may explain the incompressible states observed in the second Landau level on the flanks
of the elusive $\nu=5/2$ state.

The RR states (or clustered states) in ultracold rotating atomic
gases have been already the focus of several works. Numerically exact
diagonalizations of small systems have provided some hints of the
presence of RR states.
In the seminal work by Cooper, Wilkin and Gunn~\cite{Cooper01},
spectra in the torus geometry exhibit the special ground state
degeneracy associated with the topology of the RR states and 
have excellent overlaps
with the explicit RR trial wavefunctions. In the spherical geometry
there is also a set of incompressible states with the special relationship
between the flux and the number of particles of the RR states~\cite{Regnault03}.
Extrapolation of the gap points to a non-zero value for the MR
$\nu=1$ case, whereas the $\nu=3/2$ and $\nu=2$ results do not show clear
evidence for a smooth thermodynamic limit.  On the sphere geometry,
the overlap is excellent for the MR state and tend
to a nonzero value as the system size increases~\cite{Chiachen05}.
A more recent work~\cite{Rezayi05} has been done on the $\nu=3/2$
case. It shows that the overlap can be improved by adding a longer range
 dipole-dipole interaction 

Our purpose is to go beyond existing studies and look at size effects
for fractions $\nu=3/2,2,5/2$ using exact diagonalizations on the
sphere. We also check if the quasihole ground states are
present at these filling factors by evaluating overlap between
subspaces spanned by these states and the lowest energy excitations
of the delta function interaction, $s$-wave scattering system. 
Appearance of such quantum
states with the correct degeneracy predicted by conformal field theory arguments
is a strong hint of the validity of the RR state hypothesis.

In section \ref{parafermion}, we give an overview of the clustered
states and their excitations while section \ref{CFT} is devoted to
the conformal field theory (CFT) formulation. Section
\ref{numerical} is a brief description of the numerical method we
use. In section \ref{ground}, we give the results for
the overlap of the ground states. We discuss how the system size,
longer range or higher order n-body interaction impact on the
overlap. Section \ref{excitations} is devoted to the quasihole
excitations. In addition to the overlap values, we also give
numerical evaluation of quasihole degeneracy on sphere for
fractions $\nu=1, 3/2, 2, 5/2$ and compare them to a
formula due to Ardonne~\cite{Ardonne02} to check the validity of the conformal
field theory approach.

\section{Parafermionic states}\label{parafermion}

For the sake of simplicity, we use the disk geometry in this section.
In the symmetric gauge, the LLL one-body wave functions are
given by~:

\begin{eqnarray}
\phi_m(z)&=&\frac{1}{\sqrt{2\pi 2^m m!}} z^m e^{-\left|z\right|^2 / 4}\label{onebodydisk},
\end{eqnarray}
where $z=x+iy$ and we take the magnetic length $l_B$ to be equal
to unity. Any $N$-body wave function of particles in the LLL can be
written as a polynomial ${\cal P}$ in the particle $z_i$
coordinates~:

\begin{eqnarray}
\Psi\left(z_1,...,z_N\right)&=&{\cal P}\left(z_1,...,z_N\right)\;\;
e^{-\sum_i\left|z_i\right|^2/4l^2}.
\label{genericnbodydisk}
\end{eqnarray}
From now on, we drop the global Gaussian factor.
The $k$-type RR state is the exact zero energy ground state of the
pure $(k+1)$-body $\delta$-function interaction hamiltonian~:

\begin{eqnarray}
{\cal{H}}^{RR}_{k}&=&\sum_{i_1<...<i_{k+1}}\delta^{(2)}\left(z_{i_{1}}
- z_{i_{2}}\right)...\delta^{(2)}\left(z_{i_{k}} -
z_{i_{k+1}}\right).\label{nbodyhamiltonian}
\end{eqnarray}

The corresponding wave function can be written\cite{Cappelli01}~:
\begin{eqnarray}
\Psi^{RR}_{k}&=&\sum_{\sigma}'\prod_{0 \le r < s < N/k}
\chi(z_{\sigma(kr+1)},...,z_{\sigma(kr+k)};\nonumber\\
&&z_{\sigma(ks+1)},...,z_{\sigma(ks+k)}),\label{parawavefunction}
\end{eqnarray}
where
\begin{eqnarray}
&&\chi\left(z_1,...,z_k;z_{k+1},...,z_{2k}\right)=\left(z_1-z_{k+1}\right)\nonumber\\
&&\left(z_1-z_{k+2}\right)\left(z_2-z_{k+2}\right)\left(z_2-z_{k+3}\right)\nonumber\\
&&...\left(z_k-z_{2k}\right)\left(z_k-z_{k+1}\right).\label{chifactor}
\end{eqnarray}
The sum is over all permutations $\sigma$ of $N$ elements such
that $\sigma(1)<\sigma(k)<...<\sigma(N-k+1)$. The number of
particles $N$ must be a multiple of $k$.

These states are also referred to as clustered states because
the wavefunction 
(\ref{parawavefunction}) vanishes when $k+1$ or more particles are
at the same position. The $k$-type RR state is associated to the
filling factor $\nu=k/2$. Each RR state is the zero-energy ground
state of its corresponding Hamiltonian with the smallest total
angular momentum.

The simplest case $k=1$ corresponds to the usual Laughlin wave function~:
\begin{eqnarray}
\Psi_{Laughlin}&=&\prod_{i<j}\left(z_i-z_j\right)^2\label{laughlin}
\end{eqnarray}
and is the exact ground state for rotating bosons with $s$-wave
scattering at $\nu=1/2$ whose effective Hamiltonian is given by~:
\begin{eqnarray}
H_{\rm LLL}=g l_B^2\sum_{i<j} \delta^{(2)}\left({\bf r}_i-{\bf r}_j\right)&{\rm and}&g=\sqrt{8\pi}
\;\hbar \omega_c \frac{a_s}{l_z}.
\label{hamiltonian2DLLL}
\end{eqnarray}
where $a_s$ is the $s$-wave scattering length, $l_z$ is the
characteristic length of the ${\bf\hat{z}}$ axis oscillator which
is used for 2d confinement, and $\omega_c$ is the cyclotron
rotation frequency.

The case $k=2$ is the so called MR/Pfaffian state. It can be rewritten as~:
\begin{eqnarray}
\Psi_{Pfaffian}&=&{\rm Pf}\left(\frac{1}{z_i-z_j}\right)\prod_{i<j}\left(z_i-z_j\right) ,\label{pfaffian}
\end{eqnarray}
where ${\rm Pf}$ stands for the pfaffian defined as~:
\begin{eqnarray}
{\rm Pf}\left(A\right)&=&\sum_{\sigma} \epsilon_{\sigma} A_{\sigma(1)\sigma(2)}
A_{\sigma(3)\sigma(4)}...A_{\sigma(N-1)\sigma(N)},\label{pfaffiandefinition}
\end{eqnarray}
where $A$ is a skew-symmetric $N\times N$ matrix (N even), the sum
runs over all permutations of the index with N values and
$\epsilon_{\sigma}$ is the signature of the permutation.

If we deviate from the clustered state at filling factor $k/2$ by
adding $\Delta\Phi$ vortices (or flux quanta in the 2DES analog),
quasihole excitations are generated. For each added vortex, $k$
quasiholes are nucleated. For the Laughlin state, quasihole ground
state wave functions can easily be obtained. Any function of the form~:
\begin{eqnarray}
\Psi_{Laughlin}^{qh}&=&P\left(z_1,...,z_N\right)\Psi_{Laughlin},\label{laughlinquasihole}
\end{eqnarray}
where $P$ is a symmetric polynomial, corresponds to a zero-energy
many quasihole state. For one quasihole at position $w_1$, the
polynomial $P$ is just $\prod_i(z_i-w_1)$. Read and Rezayi have also
obtained an explicit formula in the case of the MR
state\cite{Read96} for two quasiholes at positions $w_1$ and
$w_2$~:
\begin{eqnarray}
\Psi_{Pfaffian}^{2qh}&=&{\rm Pf}\left(\frac{f\left(z_i,z_j;w_1,w_2\right)}{z_i-z_j}\right)
\prod_{i<j}\left(z_i-z_j\right) ,\label{pfaffianquasihole}
\end{eqnarray}
with $f\left(z_i,z_j;w_1,w_2\right)=
\left(z_i-w_1\right)\left(z_j-w_2\right)+\left(z_i-w_2\right)\left(z_j-w_1\right)$.
In the general case, the quasihole ground states can be written down
using the CFT formulation. This formalism also reveals their non-Abelian
statistics.

\section{CFT approach}\label{CFT}

There is an elegant way to introduce RR states involving
CFT\cite{Read99}. The key idea\cite{Moore91} is to express the
wave function as a correlator using the algebra of the
$\mathbb{Z}_k$ parafermions\cite{Zamolodchikov85}. This
algebra is defined a set of field
$\left\{\Psi_1(z),...,\Psi_{k-1}\right\}$ obeying the following
operator product expansion (OPE)~:
\begin{eqnarray}
\psi_l(z)\psi_{l'}(z') & \sim & d_{l,l'}\left(z-z'\right)^{-\left(\Delta_l
+\Delta_{l'}-\Delta_{l+l'}\right)}\nonumber\\
            & & \times \psi_{l+l'}(z') + \cdots\;\;\;(l+l'<k),\label{opepsipsi}\\
\psi_l(z)\psi_{l'}^\dagger(z') & \sim & d_{l,k-l'}\left(z-z'\right)^{-\left(\Delta_l+\Delta_{l'}
-\Delta_{l-l'}\right)}\nonumber\\
        & &     \times \psi_{l-l'}(z') + \cdots\;\;\;(l'<l),\label{opepsipsid}\\
\psi_l(z)\psi_{l}^\dagger(z') & \sim & \left(z-z'\right)^{-2\Delta_l}\nonumber\\
        & &     \times \left(\mathbb{I} + \frac{2\Delta_l}{c}\left(z-z'\right)^2 T(z')
        + \cdots\right),\label{opepsipsim}\\
T(z)\psi_l(z')&\sim&\frac{\Delta_l}{\left(z-z'\right)^2}\psi_l(z')\nonumber\\
        & &     +\frac{1}{z-z'}\partial\psi_l(z')+\cdots.\label{opetpsi}
\end{eqnarray}
where $\psi_l^\dagger=\psi_{k-l}$, $T(z)$ is the stress-energy tensor,
$\Delta_l$ is
the conformal weight  of the field $\psi_l$, $c$ is
the theory central charge and $d_{l,l'}$ are numerical
coefficients. The algebra of $\mathbb{Z}_k$ parafermions
corresponds to the choice $\Delta_l=l(k-l)/k$ leading to the
central charge $c=2(k-1)/(k+2)$ and to uniquely determined
$d_{l,l'}$ coefficients.

Read and Rezayi have shown that the following wave function~:
\begin{eqnarray}
\Psi^{RR,CFT}_{k}&=&\left\langle \psi_1\left(z_1\right)\cdots\psi_1
\left(z_N\right)\right\rangle\;\prod_{i<j}\left(z_i-z_j\right)^{2/k},\label{parawavefunctionCFT}
\end{eqnarray}
 is equivalent to expression \pref{parawavefunction}. 
One can easily show that this expression vanishes quadratically as $k+1$ particles 
go to the same point using the OPE rules above \prefs{opepsipsi}{opetpsi}. Within this formalism, 
it has been argued that the zero energy quasihole states can be built by inserting a spin field 
for each quasihole into the correlator of \pref{parawavefunctionCFT}. For $k=2$, this 
spin field $\sigma$ is equivalent to the magnetization operator of the Ising model. 
In the Ising case, the fusion rules are given by~:
\begin{eqnarray}
\sigma(z)\psi_{1}(z') &=& \frac{1}{\left(z-z'\right)^{1/2}} \psi_{1}(z'),\label{opesigmapsi}\\
\sigma(z)\sigma(z') &=&  \frac{1}{\left(z-z'\right)^{1/8}} \mathbb{I} +  \left(z-z'\right)^{3/8}
\psi_{1}(z').\label{opesigmasigma}
\end{eqnarray}
For $n=2\Delta\Phi$ quasiholes, the candidate state is then~:
\begin{eqnarray}
\Psi^{n\, qh}_{Pfaffian}&=&\left\langle \psi_1\left(z_1\right)...\psi_1
\left(z_N\right)\sigma\left(w_1\right)...\sigma\left(w_n\right)\right\rangle\nonumber\\
&\times& \prod_{i<j}\left(z_i-z_j\right)\;\;\prod_{i}\prod_{p=1}^{n}\left(z_i-w_p\right)^{1/2}.
\label{quasiholePfaffianCFT}
\end{eqnarray}
The fusion rule \pref{opesigmasigma} leads to a non-trivial
degeneracy of the quasihole states~: there are $2^{n/2-1}$ ways to
fuse the spin operators leading to a non zero correlator, thus
giving as many different wavefunctions. This so-called
intrinsic degeneracy is the key of non-Abelian statistics~:
exchanging two quasihole coordinates of a given quasihole state
will result in a linear combination of states of the same family
instead of an overall multiplicative phase factor. In the case of
the spherical geometry that we will discuss later, an additional
(extrinsic) degeneracy arise from the Laughlin-like part of
Eq.\pref{quasiholeCFT}. Determining the multiplet decomposition of
quasihole states in such a case is a challenging 
task~\cite{Read96, Gurarie00, Ardonne02} and constitutes a non-trivial
check of the CFT approach when compared to numerical calculations.
More details will be given in section \ref{excitations}.

For $k>2$, the spin field that we have introduced has to be replaced by
one of the primary field operators of the $\mathbb{Z}_k$
parafermion algebra. The guess is to use the operator $\sigma_1$ which minimizes the charge of the
quasiholes. The wavefunction \pref{quasiholePfaffianCFT} can then be
generalized to~:
\begin{eqnarray}
\Psi^{n\, qh}_{k}&=&\left\langle \psi_1\left(z_1\right)...\psi_1
\left(z_N\right)\sigma_1\left(w_1\right)...\sigma_1\left(w_n\right)\right\rangle\nonumber\\
&\times& \prod_{i<j}\left(z_i-z_j\right)^{2/k}\;\;\prod_{i}\prod_{p=1}^{n}\left(z_i-w_p\right)^{1/k}.\label{quasiholeCFT}
\end{eqnarray}
The fusion rules involving $\sigma_1$ are more 
complex~\cite{Zamolodchikov85, Gepner87} but the same remarks as for the
$k=2$ case apply, meaning they lead to non-Abelian statistics.
Notice that for $k\ge 3$, such states have been proposed to be a
robust way to implement quantum computation\cite{Freedman03}.

\section{Numerical method}\label{numerical}

We use exact diagonalizations to study if the RR states are
relevant to the physics of the fast rotating boson gases
at filling factor $\nu=k/2$. Numerical calculations can be done on
various geometry such as the disk, the torus or the sphere. The
disk geometry is plagued by edge effects and thus closed
geometries are preferred when dealing with bulk properties. In
this paper, all calculations are done on the spherical geometry
\cite{Fano86}. Due to the $SU(2)$ symmetry, states can be
classified with respect to their total angular momentum $L$ and
its projection along one axis $L_z$. Solutions of the one-body
problem are given by the monopole harmonics~\cite{Wu76} (a
generalization of the spherical harmonics), which take the
following form in the LLL~:
\begin{eqnarray}
Y_m(u,v)&=&\sqrt{\frac{\left(2S+1\right)!}{4\pi\left(S-m\right)!\left(S+m\right)!}}\;\;
u^{S+m}v^{S-m},
\label{harmonic}
\end{eqnarray}
where $m\hbar$ is the projection of the angular momentum, $-S\leq m\leq +S$, $u$ and
$v$ are the spinor components in spherical coordinates~:
\begin{eqnarray}
u=\cos\left(\theta/2\right)e^{i\phi/2},&&v=\sin\left(\theta/2\right)e^{-i\phi/2}.
\label{spinor:definition}
\end{eqnarray}
The radius $R$ of the sphere is related to the number of vortices
(or flux quanta in the 2DES language) $N_\phi$ that pierce
it~:
\begin{eqnarray}
R&=&l_B\sqrt{N_\phi/2}.
\label{radiusflux}
\end{eqnarray}
The one particle angular momentum $S$ is such that $2S=N_\phi$.
Due to sphere topology, the relation between the number of
particles $N$ and $N_\phi$ for a given fraction is linear with a
non-zero shift. For each trial wavefunction for a given
fraction, there is a unique shift, which is a characteristic of 
the quantum Hall state. In
the case of parafermionic states, the relation between the
magnetic flux and the number of particles is given by~:
\begin{eqnarray}
N_\phi&=\frac{2}{k} N - 2 \label{Nfluxrelation}
\end{eqnarray}
This can be deduced from the expression of the
(\ref{parawavefunction}) on the sphere by applying a
stereographic projection. Formally, we just have to drop the
gaussian factor and make the substitution~:
\begin{eqnarray}
\left(z_i-z_j\right)&\longrightarrow&\left(u_i v_j - u_j v_i\right).\label{disktosphere}
\end{eqnarray}

The two-body interaction is completely characterized by a set of
$2S+1$ numbers $\left\{V_m\right\}$ called the
pseudo-potentials~\cite{Haldane83}. The integer $m$ is the
\textit{relative} angular momentum between the two particles. For
spinless bosons, only even-$m$ potentials are relevant. $s$-wave
scattering interaction corresponds to the case where all
pseudo-potentials are equal to zero except $V_0$. Longer range
interactions involve additional pseudopotentials, the next
one for spinless bosons being $V_2$. Thus, adding some $V_2$
component allows to test the effect of longer range interactions.
Comparison between the RR states or their quasihole excitations
with the true ground states is achieved by computing overlaps. For
two states $|\Psi\rangle$ and  $|\Phi\rangle$, the overlap is
defined as ${\cal O}=\left|\left\langle \Phi | \Psi
\right\rangle\right|^2$. This definition can be extended to the
case of subspaces of same dimension $N$ and spanned by vector sets
$\{|\Phi_i\rangle\}$ and $\{|\Psi_j\rangle\}$~:
\begin{eqnarray}
{\cal O}&=&\frac{1}{N}\sum_{i,j=1}^N \left|\left\langle \Phi_i
|\Psi_j \right\rangle\right|^2.\label{overlap:definition}
\end{eqnarray}

Both exact ground states and RR/quasihole candidate ground states are
evaluated using exact diagonalizations of the associated
Hamiltonian. Numerical diagonalizations are achieved using
L\'anczos-like algorithm or full diagonalization algorithm in a
given $L_z$ subspace. In the case of $k+1$-body hardcore
interaction, the matrix are being less and less sparse with
increasing $k$ value, requiring more memory and CPU time and
making convergence harder to reach. Moreover due to the $L_z$-only
restriction on the Hilbert space, looking at the quasihole ground
states require the evaluation of highly degenerate eigenstates. Thus we
can reach lower system sizes compared to the ground state.

\section{Ground state overlaps}\label{ground}

We look at the overlap between the RR ground state and the exact
ground state. Tables \ref{tableground1} to \ref{tableground52}
display the overlaps for various fractions between the RR state and
the two-body hardcore interaction hamiltonian ground state for
different sizes. We also include overlaps with other ground
states such as Coulomb interaction or $n\geq 2$-body hardcore
interactions. In the particular case where $N=4\nu$, the overlap
is equal to one. This is due to the dimension of the Hilbert
subspace in the $L=0$ sector when $S=1$ which is equal to one.

Some of the results presented in table \ref{tableground1} have already been
published\cite{Chiachen05}. They show that the Pfaffian state 
is a good description of the
physics at $\nu=1$. As already noticed, longer range interactions
tend to improve the overlap.

For the $\nu=3/2, 2, 5/2$ fractions, the situation is not so
clear. Fewer values can be obtained and the overlap is
non-monotonic with respect to the size of system, making dubious
convergence to the thermodynamic limit. If we
consider long-range interaction like Coulomb
interaction, overlaps are improved, but we still get the same
non-monotonic behavior. The same remarks are valid for the comparison
with the $n$-body hardcore interaction $(2\leq n \leq k)$~: the
overlaps are closer to unity as $n$ tends towards $k+1$.

\begin{figure}[!htbp]
\begin{center}
\includegraphics[width=8cm, keepaspectratio]{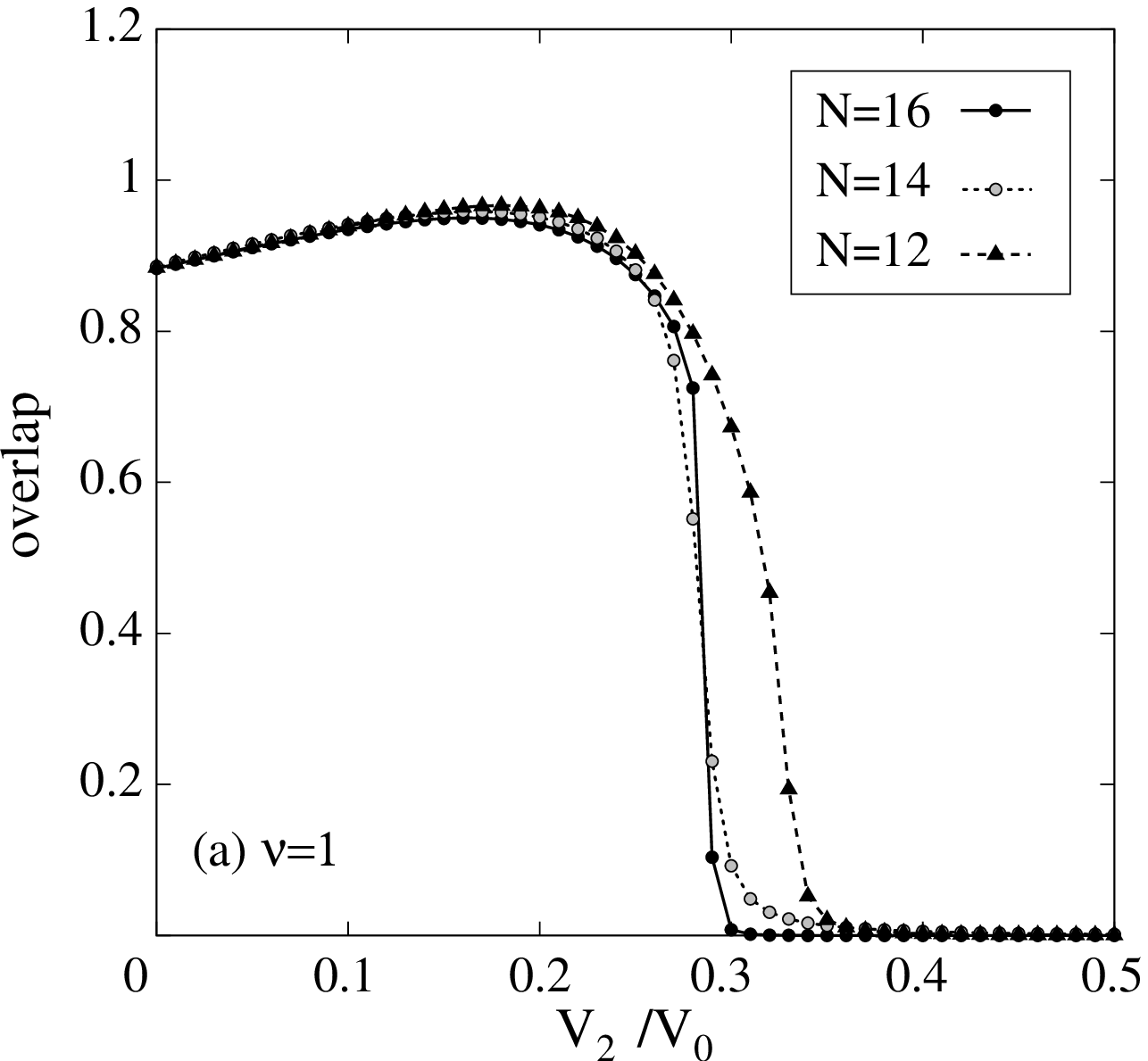}
\includegraphics[width=8cm, keepaspectratio]{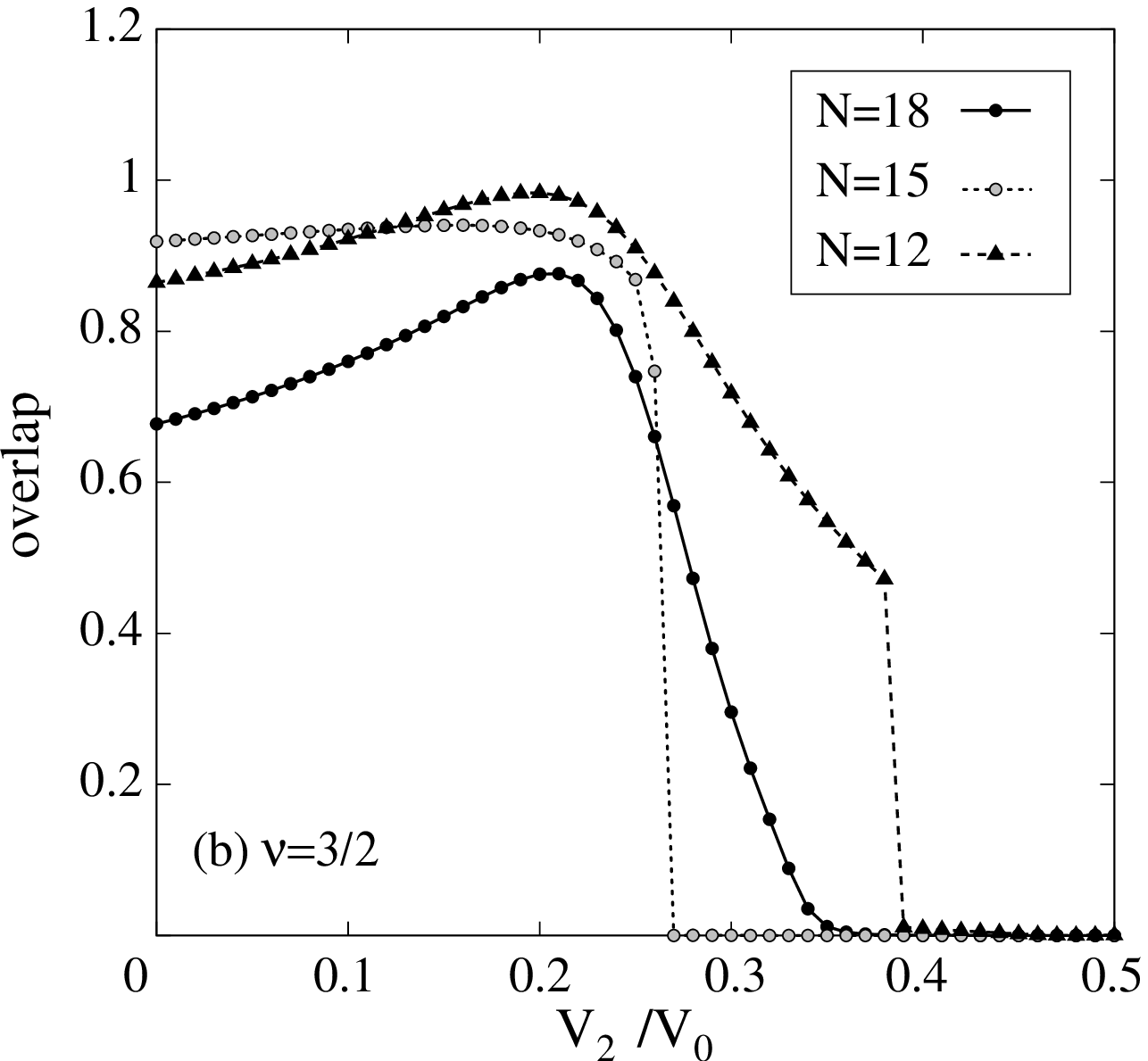}
\end{center}
\begin{center}
\includegraphics[width=8cm, keepaspectratio]{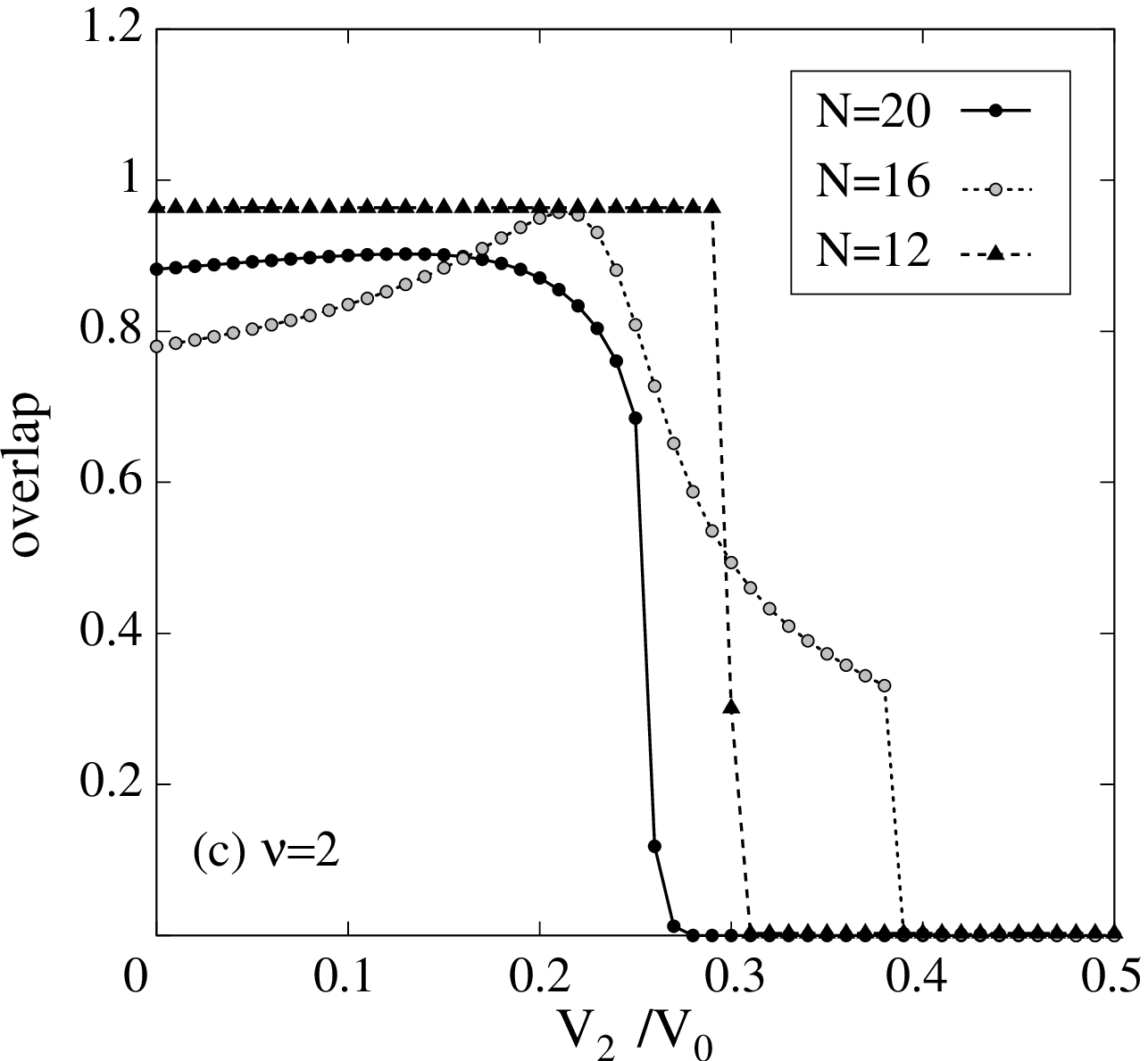}
\includegraphics[width=8cm, keepaspectratio]{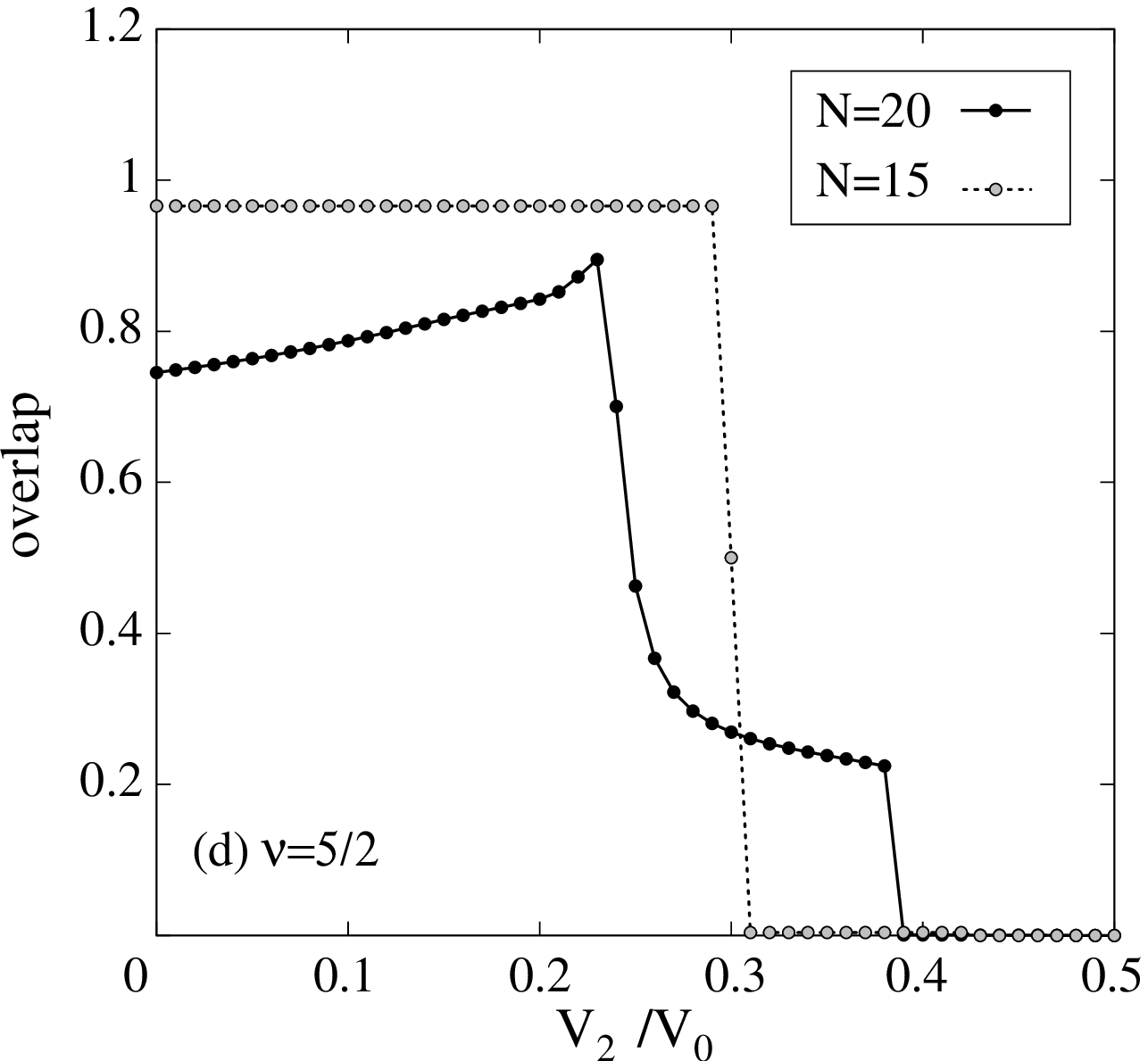}
\end{center}
\caption{From left to right and top to bottom: overlap between the
RR state and the ground state of longer range interaction
Hamiltonian as a function of $V_2/V_0$ at filling factors $\nu=1,
3/2, 2$ and $5/2$.} \label{OverlapV2V0}
\end{figure}

\begin{figure}[!htbp]
\begin{center}
\includegraphics[width=8cm, keepaspectratio]{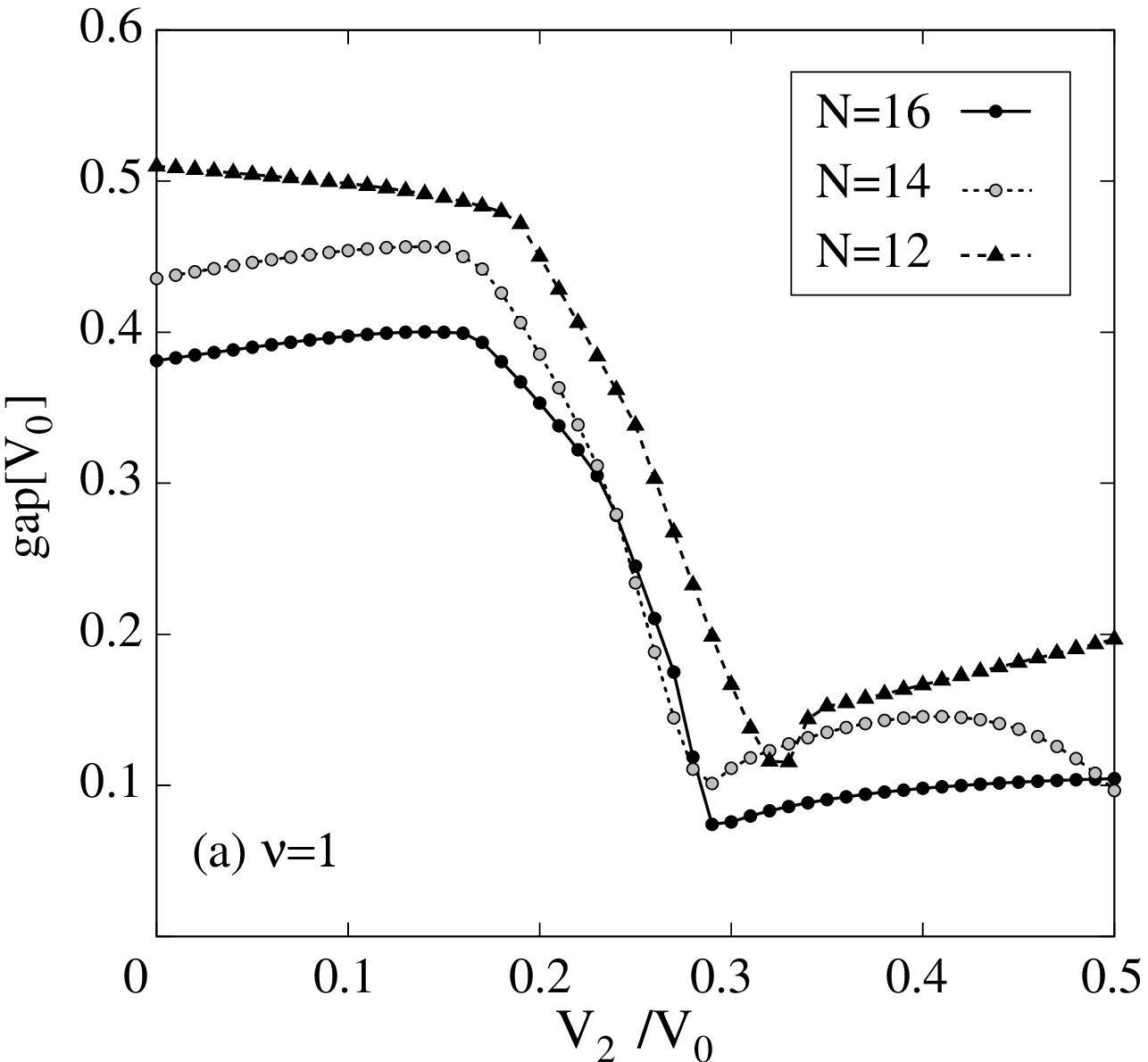}
\includegraphics[width=8cm, keepaspectratio]{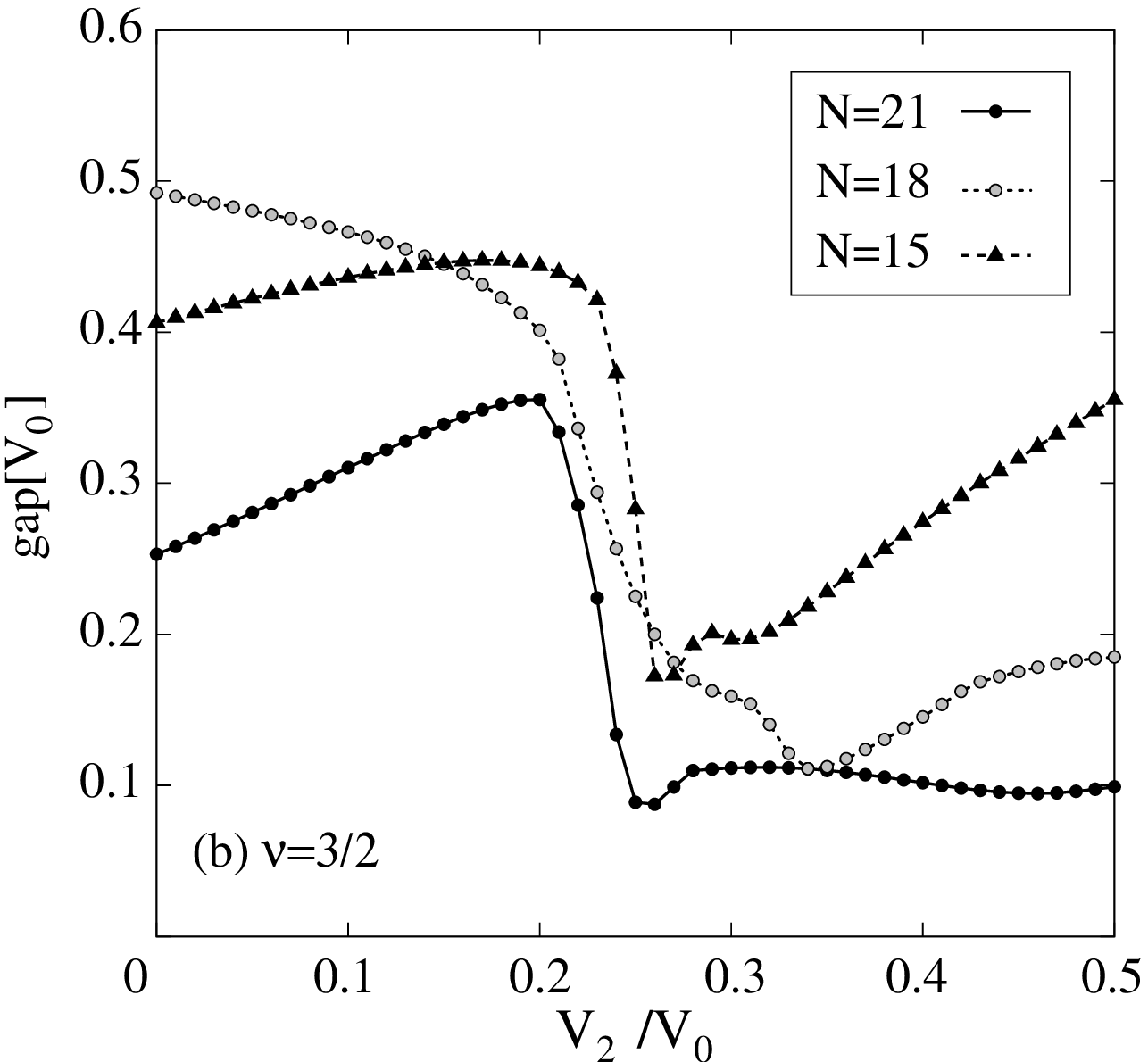}
\end{center}
\caption{Gap of the longer range interaction Hamiltonian as a
function of $V_2/V_0$ at filling factors $\nu=1$ (a) and
$\nu=3/2$ (b).} \label{GapV2V0}
\end{figure}

To ascertain the role of longer-range interaction, we
follow the method proposed in ref.(\onlinecite{Rezayi05}) for $\nu=3/2$. We
add a $V_2$ contribution to the two-body hardcore interaction.
Figure \ref{OverlapV2V0} shows the overlaps as a function of the ratio
$V_2/V_0$ for the four filling factors $\nu=1,3/2,2$ and $5/2$.
The conclusions we can draw are similar to ref.(\onlinecite{Rezayi05})~:
long-range interactions help to
stabilize the parafermionic ground state. Note that the drop of
the overlap for large values of $V_2/V_0$ is correlated to a
similar effect in the gap value (see figure \ref{GapV2V0}) and is
thus related to the loss of incompressibility.

\section{Quasihole excitations}\label{excitations}

If we believe that the parafermionic description is relevant for the fractions
$\nu=k/2$, then quasihole excitations should also be
present. Studying quasihole excitation on the sphere geometry is
 interesting on its own. Indeed,
non-Abelian statistics is related to the quasihole ground state
degeneracy. We can sort these states by their orbital quantum
numbers $L$ and $L_z$. Evaluating the degeneracy of each sector
is already a non trivial task. A formula was found for the Pfaffian case
by Read and Rezayi~\cite{Read96}. Gurarie and Rezayi~\cite{Gurarie00} 
have also an algorithm to compute the degeneracy in
the $\nu=3/2$ case. Finally, Ardonne has proposed an expression for
the degeneracy valid for any $\nu=k/2$ value. We briefly describe how we
extract the multiplet decomposition of the quasihole
degenerate states from Ardonne's formula in an Appendix.
Comparison of degeneracy values obtained using the CFT approach with the results
of numerical exact diagonalizations is a way to
validate the CFT approach. Numerical computations have been
performed for the Pfaffian~\cite{Read96} at $\nu=1$ and
also~\cite{Gurarie00} for the $\nu=3/2$ case. We give here additional values for
these two fractions (tables \ref{tabledegeneracy1} and
\ref{tabledegeneracy32}). We also compute degeneracies for $\nu=2$
and $\nu=5/2$ which haven't been published before (see tables
\ref{tabledegeneracy2} and \ref{tabledegeneracy52}). The results
we obtain are in agreement with Ardonne's formula.

To test the validity of the quasihole hypothesis, we compute the
overlap at a given fraction $\nu=k/2$ and for $k$ quasiholes
between the subspace spanned by the quasihole states of the
$k+1$-body hardcore Hamiltonian (\ref{nbodyhamiltonian}) and the
lowest energy states of the short-range problem at each $L$ value. In
each $(L,L_z)$ sector, we thus consider the $N_L^{k,q}$ lowest energy
eigenstates where $N_L^{k,q}$ is the degeneracy for $q$ quasihole
ground states at filling factor $\nu=k/2$ with angular momentum
$L$ ($N_L^{k,q}$ is $L_z$ independent) as candidates for the non-Abelian quasihole states. The corresponding overlap
${\cal O}_L^{k,q}$ is evaluated using definition Eq.(\ref{overlap:definition}).
In order to easily characterize the agreement with the whole set of
quasihole states for all $L$ values, we introduce a
total overlap defined as~:
\begin{eqnarray}
{\cal O}^{k,q}&=&\frac{\sum_L N_L^{k,q}\left(2L+1\right){\cal O}_L^{k,q}}{\sum_L N_L^{k,q}\left(2L+1\right)},\label{overlap:total}
\end{eqnarray}
which is just another way to write the total overlap with respect
to the subspace spanned by all quasihole states.

\begin{figure}[!htbp]
\begin{center}
\includegraphics[width=8cm, keepaspectratio]{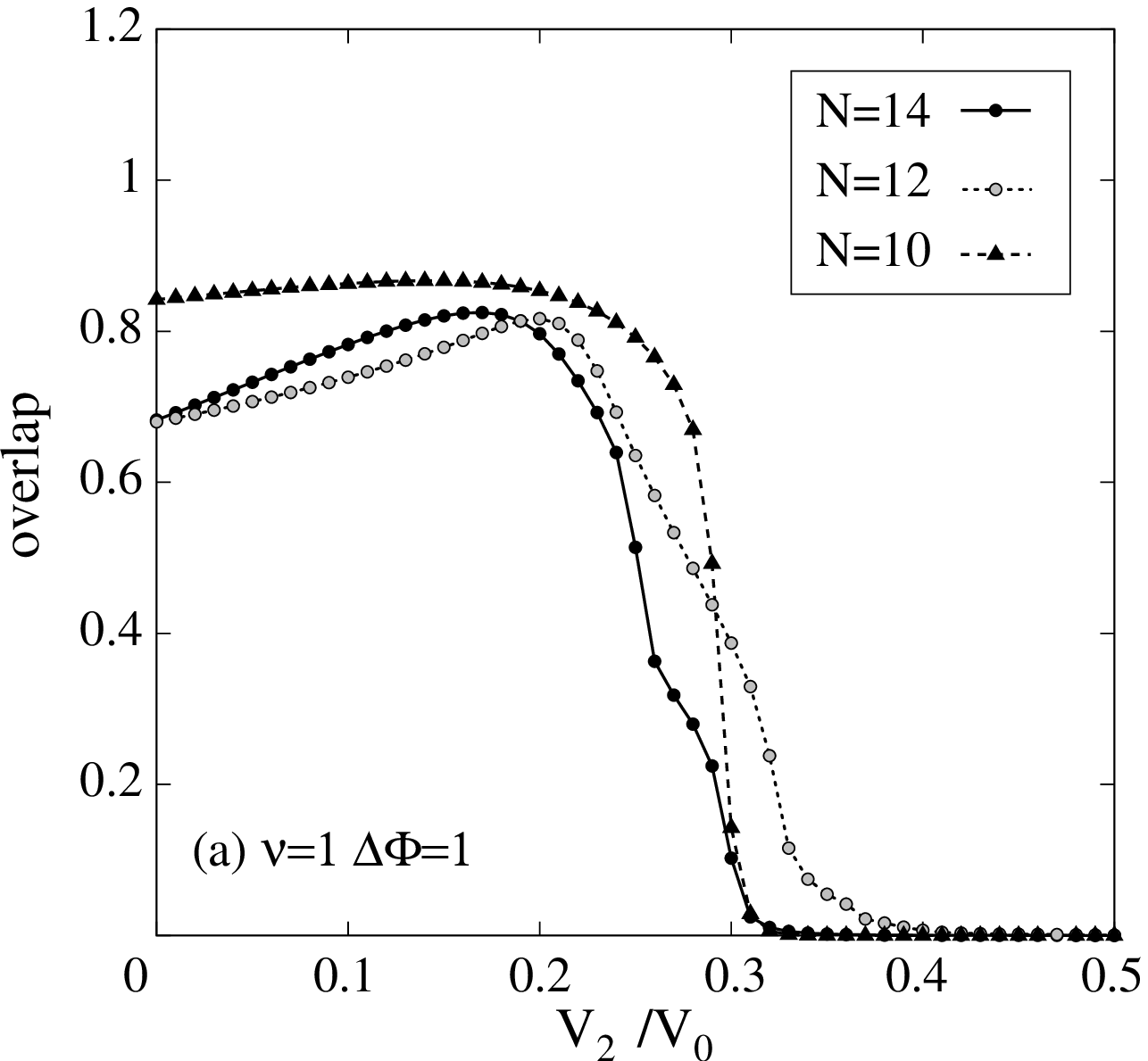}
\includegraphics[width=8cm, keepaspectratio]{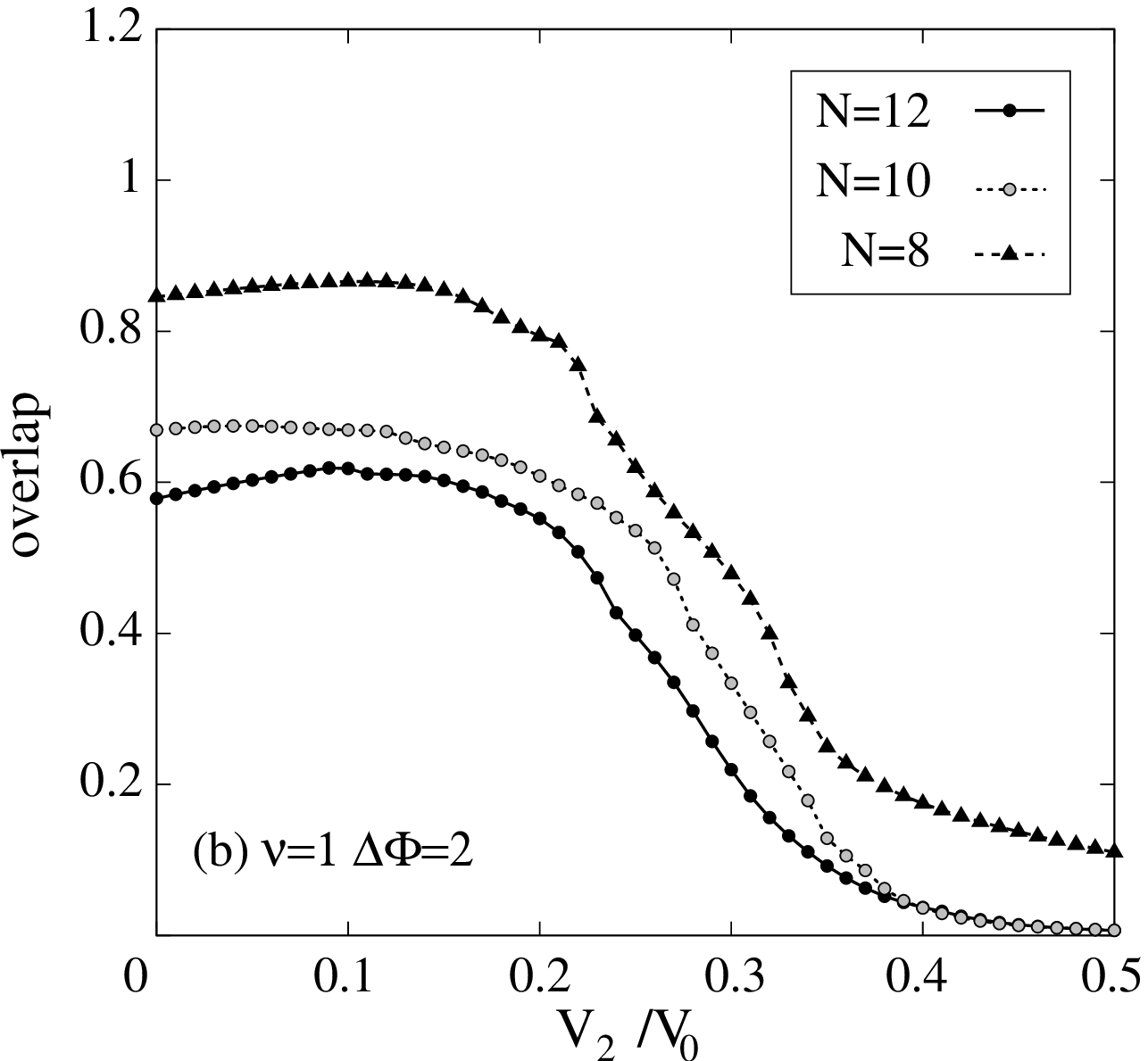}
\end{center}
\begin{center}
\includegraphics[width=8cm, keepaspectratio]{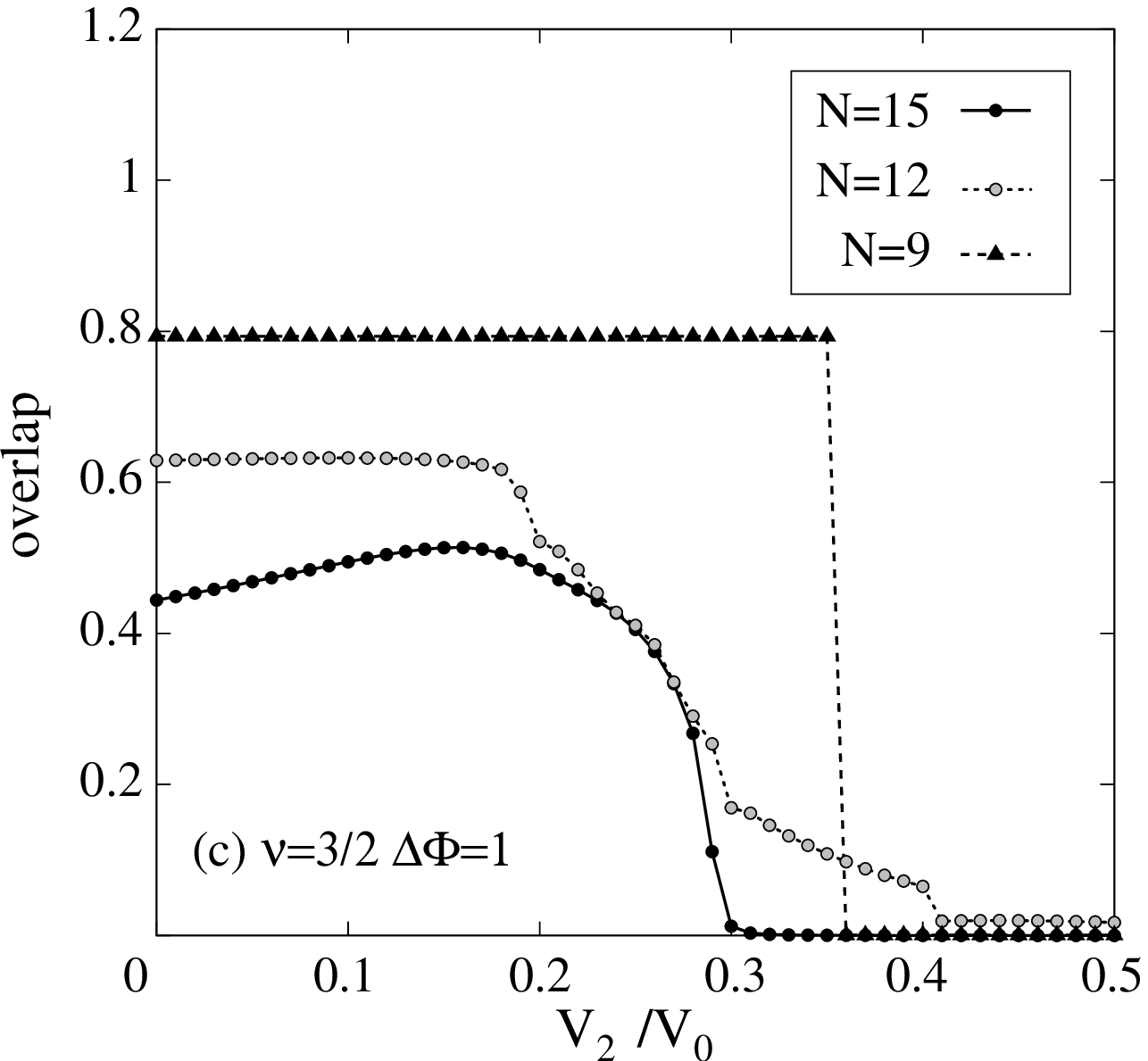}
\includegraphics[width=8cm, keepaspectratio]{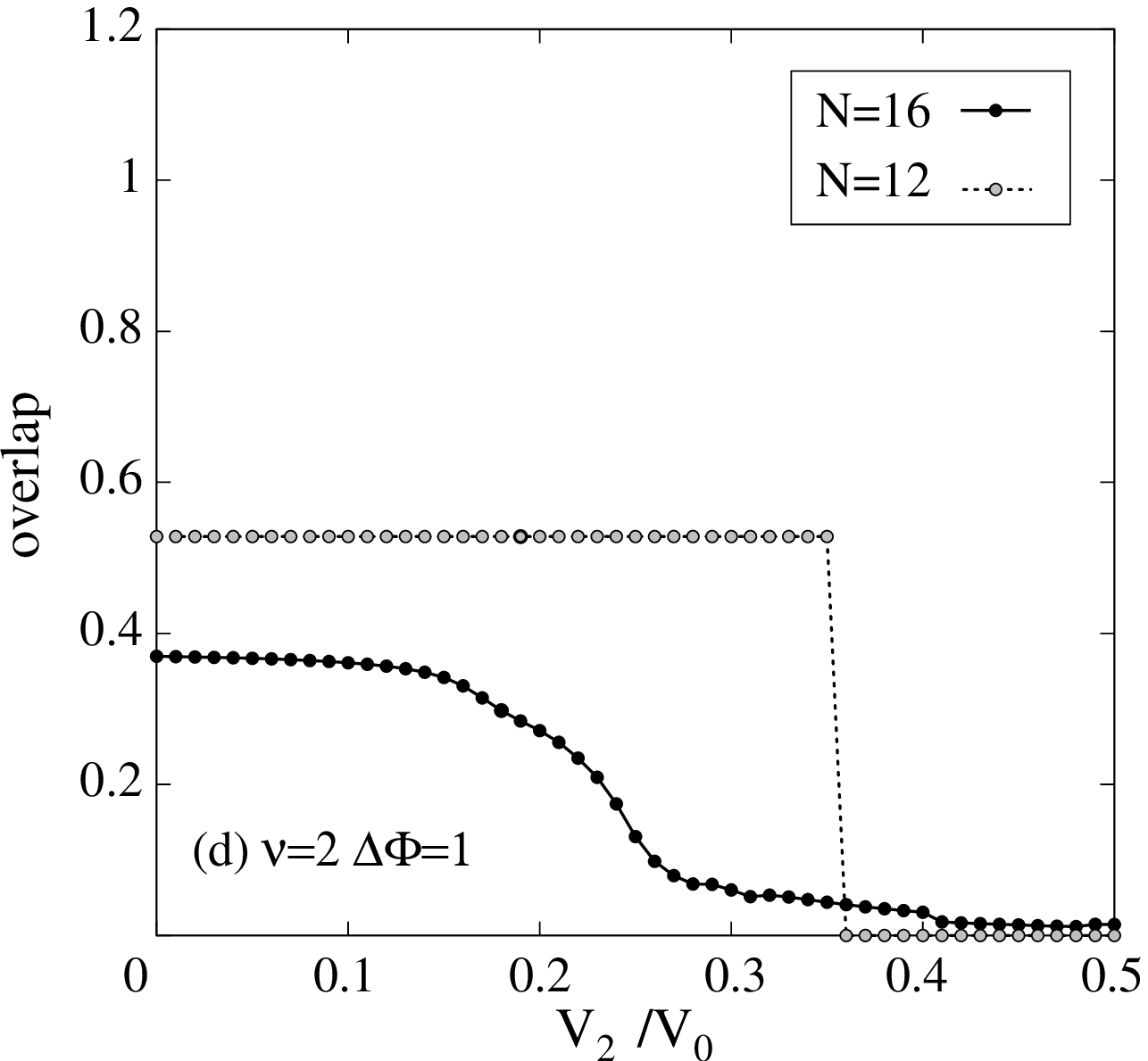}
\end{center}
\caption{Upper part: total overlap for as a function of $V_2/V_0$
at filling factors $\nu=1$ with two quasiholes (a,
$\Delta\Phi=1$) and four quasiholes (b, $\Delta\Phi=2$).
Lower part: total overlap as a function of $V_2/V_0$ at filling
factors $\nu=3/2$ (c) and $\nu=2$ (d) for
$\Delta\Phi=1$.} \label{OverlapQuasiholeV2V0}
\end{figure}

Our results are given in tables \ref{tableoverlap1},
\ref{tableoverlap32}, \ref{tableoverlap2} and \ref{tableoverlap52}
for fractions $\nu=1,3/2,2$ and $5/2$ and $q=k$ quasiholes.
The success of  the quasihole description is quite impressive at $\nu=1$.
However the agreement becomes increasingly worse with higher $k$ values.
Notice that for a given system with fixed $k$ and $N$ values, the
smallest overlap is obtained for the largest $L$ total momentum.
Due to its high $L_z$ degeneracy, it plagues the total overlap.
This certainly means that fewer quasihole excitation with
non-Abelian statistics are present in the pure hard-core model than in the
$k+1$-body system. We can also add more quasiholes. Table
\ref{tableoverlap12} displays the results for $q=2k$ at $\nu=1$.
Considering the high degeneracy we are looking at (up to 336 for
$N=12$), the overlaps are quite good especially if we do not take
into account the ones associated to the largest total momentum.
The effect of longer range interaction is similar to the ground
state case~: adding some $V_2$ component tends to improve the
overlap. In figure \ref{OverlapQuasiholeV2V0}, we have plotted the
total overlap as a function of $V_2/V_0$ for fractions $\nu =1,3/2,2$.
There is now a maximum of the overlap for a moderate amount of longer-range
interactions.

\section{Conclusion}\label{conclusion}

Existence of quasiparticles with non-Abelian statistics is an
exciting question of modern physics. The possible appearance of
such quasiparticles in rotating ultracold boson gases is a strong
motivation to experimentally reach the corresponding regime. 
At filling factor $\nu=1$, we have shown that the pairing scheme of Moore and Read
extends to the quasihole excitations. The degeneracy we observe 
is exactly that predicted by the CFT approach and the overlap of 
the subspaces spanned by the quasiholes shows that they are likely 
to be relevant at this fraction. Concerning the RR states for larger filling factors
their overlaps with the ground state of the pure hard-core model
are much less impressive and the size dependence is irregular,
as was already observed in the gap values. We have found a set of degenerate states
with the quantum numbers predicted for quasiholes generated from the
RR states by addition of flux quanta. they have also the features
expected from the CFT approach. Again the overlaps (now for subspaces
taken as a whole) are much less impressive. It is highly unlikely that 
the RR states are relevant for large $k$ values. Adding long-range
interactions like a second pseudopotential $V_2$ certainly strengthen
these RR states and their quasiholes states. So to construct
in practice a RR state one may have to fine-tune the interaction potential between
ultracold atoms. It is likely however that the Pfaffian at $\nu =1$
is the most conveniently implemented state for manipulation of non-Abelian
statistics (but it does not support universal quantum 
computation~\cite{Freedman03}).

\section{Acknowledgment}

We thank Chiachen Chang and Jainendra Jain for useful
discussions. We thank IDRIS-CNRS for a computer time
allocation.

\section{Apppendix}

Our purpose is to show how we can get the multiplet degeneracy for
the quasihole states from Ardonne's formula. The formula
gives access to the intrinsic degeneracy and is derived from the
truncated characters of the $\mathbb{Z}_k$ algebra (or
$\mathfrak{su}(2)_k/\mathfrak{u}(1)$). These truncated characters
can be written as~:

\begin{eqnarray}
Y_n\left(x;q,k\right)&=&\sum_{a_i}q^{\frac{1}{2}{\bf a}.\mathbb{C}_{k-1}.{\bf a}}x^{\sum_{i=1}^{k-1} ia_i}\nonumber\\
    &\times&\prod_{i=1}^{k-1}\left[\begin{array}{c}\frac{in}{k}+\left(\left(\mathbb{I}_{k-1}-\mathbb{C}_{k-1}\right).{\bf a}\right)_i\\a_i\end{array}\right],\label{ArdonneFormula}
\end{eqnarray}
where the q-deformed binomial is defined as follow~:
\begin{eqnarray}
\left[\begin{array}{c}m\\p\end{array}\right]&=&\frac{\prod_{i=1}^m{\left(1-q^i\right)}}{\prod_{i=1}^p{\left(1-q^i\right)}\prod_{i=1}^{m-p}{\left(1-q^i\right)}},\label{QDeformedBinomial}
\end{eqnarray}
and is equal to zero if $p>m$ or $m,p < 0$. $n$ is the number of
quasiholes. It is linked to the number of added quantum fluxes
$\Delta\Phi$ by the relation $n=k\Delta\Phi$. $\mathbb{I}_{k-1}$
is the identity dimensional matrix and
$\mathbb{C}_{k-1}=2\mathbb{A}_{k-1}^{-1}$ where $\mathbb{A}_{k-1}$
is the Cartan matrix of the $su(k)$ algebra~:

\begin{eqnarray}
(\mathbb{A}_{k-1})_{i,j}&=&2\delta_{i,j}-\delta_{|i-j|,1}.\label{Cartansuk}
\end{eqnarray}

${\bf a}=(a_1,...,a_{k-1})$ is a vector of $k-1$ non-negative
integer such that $\sum_i i a_i=F$ where $F$ is a multiple of $k$.
When we look at a system of $N$ bosons, we are only interested in
the values $F=0,k,2k,...,N$. $F$ has to be understood as the
number of unclustered bosons. In the simplest case $k=2$, $N-F$
corresponds to the number of parafermionic fields that appear when
using OPE. Each $q$-polynomial in front of given $x^F$ monomial,
is associated to the multiplet decomposition of the intrinsic
degeneracy of a given $F$ value. It will be of the form~:
\begin{eqnarray}
&\sum_L n_L \sum_{L_z=-L}^{L} q^{\alpha+L_z},&\label{qpolynomial}
\end{eqnarray}
where $\alpha$ is a global shift of the $q$ power and $n_l$ is the
number of multiplet of momentum $l$. Thus the multiplet
decomposition of the intrinsic degeneracy can be directly read-out
from the polynomial expression.

For example, let evaluate (\ref{ArdonneFormula}) for $k=3,n=6$,
For the $N=6$ bosons case, we only need the partial development
given in (\ref{ArdonneFormulaExample}) up to $x^6$~:
\begin{eqnarray}
Y_6\left(x;q,3\right)&=&1+x^3\left(q^4+q^3+q^2\right)+x^6q^6+...\label{ArdonneFormulaExample}
\end{eqnarray}
Thus the multiplet decomposition for the intrinsic degeneracy is
one singlet $L=0$ for $F=0$, one multiplet $L=1$ for $F=3$ and one
singlet $L=0$ for $F=6$. The extrinsic part gives the following
multiplet: one singlet $L=0$ for $F=6$, one multiplet $L=3$ for
$F=3$, one multiplet for $L=0, 2, 3, 4, 6$ for $F=0$. Using the
standard momentum addition rules, we get the result display in
table \ref{tabledegeneracy32}.

\begin{table*}
\begin{ruledtabular}
\begin{tabular}{ccc}
$N$ & ${\cal O}_{\text{k=1}}$ & ${\cal O}_{1/r}$\\  \hline
4 & 1.0 & 1.0\\
6 & 0.9728 & 0.9728\\
8 & 0.9669 & 0.9771\\
10 & 0.9592 & 0.9659\\
12 & 0.8844 & 0.9165\\
14 & 0.8858 & 0.9213\\
16 & 0.8833 & 0.9170
\end{tabular}
\end{ruledtabular}
\caption{Overlaps at $\nu = 1$ between the Pfaffian obtained as the ground state of the
$3$-body hardcore interaction hamiltonian and the ground state of
the $2$-body hardcore interaction or the Coulomb interaction
$(1/r)$ hamiltonian.} \label{tableground1}
\end{table*}

\begin{table*}
\begin{ruledtabular}
\begin{tabular}{cccc}
$N$ & ${\cal O}_{k=1}$ & ${\cal O}_{k=2}$ & ${\cal O}_{1/r}$\\  \hline
6 & 1.0 & 1.0 & 1.0\\
9 & 0.9642 & 0.9891 & 0.9642\\
12 & 0.8647 & 0.9702 & 0.8904\\
15 & 0.9189 & 0.9788 & 0.9307\\
18 & 0.6774 & 0.9239 & 0.7226
\end{tabular}
\end{ruledtabular}
\caption{Overlaps at $\nu = 3/2$ between the RR state obtained as the ground state of the
$4$-body hardcore interaction hamiltonian and the ground state of
the $k+1$-body hardcore interaction or the Coulomb interaction
$(1/r)$ hamiltonian.} \label{tableground32}
\end{table*}

\begin{table*}
\begin{ruledtabular}
\begin{tabular}{ccccc}
$N$ & ${\cal O}_{k=1}$ & ${\cal O}_{k=2}$ & ${\cal O}_{k=3}$ & ${\cal O}_{1/r}$\\  \hline
8 & 1.0 & 1.0 & 1.0  & 1.0\\
12 & 0.9636 & 0.9811 & 0.9949 & 0.9636\\
16 & 0.7801 & 0.8919 & 0.9753 & 0.8037\\
20 & 0.8822 & 0.9499 & 0.9874 & 0.8985
\end{tabular}
\end{ruledtabular}
\caption{Overlaps at $\nu = 2$ between the RR state obtained as the ground state of the
$5$-body hardcore interaction hamiltonian and the ground state of
the $k+1$-body hardcore interaction or the Coulomb interaction
$(1/r)$ hamiltonian.} \label{tableground2}
\end{table*}

\begin{table*}
\begin{ruledtabular}
\begin{tabular}{cccccc}
$N$ & ${\cal O}_{k=1}$ & ${\cal O}_{k=2}$ & ${\cal O}_{k=3}$ & ${\cal O}_{k=4}$ & ${\cal O}_{1/r}$\\  \hline
10 & 1.0 & 1.0 & 1.0 & 1.0  & 1.0\\
15 & 0.9659 & 0.9789 & 0.9902 & 0.9975 & 0.9659\\
20 & 0.7455 & 0.8291 & 0.9165 & 0.9798 & 0.7644
\end{tabular}
\end{ruledtabular}
\caption{Overlaps at $\nu = 5/2$ between the RR state obtained as the ground state of the
$6$-body hardcore interaction hamiltonian and the ground state of
the $k+1$-body hardcore interaction or the Coulomb interaction
$(1/r)$ hamiltonian.} \label{tableground52}
\end{table*}

\begin{table*}
\begin{ruledtabular}
\begin{tabular}{cccccccccccccccccccccccc}
$N$ & $\Delta\Phi$ & $\#$ & $L=0$ & $1$ & $2$ & $3$ & $4$ & $5$ & $6$ & $7$ & $8$ & $9$ & $10$ & $11$ & $12$ & $13$ & $14$  & $15$ & $16$ & $17$ & $18$ & $19$ & $20$\\  \hline
$4$ & $1$ & $6$ & $1$  & $0$ & $1$  &  &  &   &   &   &   &   &   &   &   &   &   &   &   &   &  &  & \\
$4$ & $2$ & $20$ & $1$ & $0$ & $2$ & $0$ & $1$ &  &  &  &  &  &  &  &  &  &  &  &  &  &  &  & \\
$4$ & $3$ & $49$ & $1$ & $0$ & $2$ & $1$ & $2$ & $0$ & $1$ &  &  &  &  &  &  &  &  &  &  &  &  &  & \\
$4$ & $4$ & $100$ & $2$ & $0$ & $2$ & $1$ & $3$ & $1$ & $2$ & $0$ & $1$ &  &  &  &  &  &  &  &  &  &  &  & \\
$6$ & $1$ & $10$ & $0$ & $1$ & $0$ & $1$ &  &  &  &  &  &  &  &  &  &  &  &  &  &  &  &  & \\
$6$ & $2$ & $50$ & $2$ & $0$ & $2$ & $1$ & $2$ & $0$ & $1$ &  &  &  &  &  &  &  &  &  &  &  &  &  & \\
$6$ & $3$ & $168$ & $0$ & $3$ & $1$ & $4$ & $2$ & $3$ & $2$ & $2$ & $0$ & $1$ &  &  &  &  &  &  &  &  &  &  & \\
$6$ & $4$ & $444$ & $3$ & $1$ & $5$ & $3$ & $7$ & $3$ & $6$ & $3$ & $4$ & $2$ & $2$ & $0$ & $1$ &  &  &  &  &  &  &  & \\
$8$ & $1$ & $15$ & $1$ & $0$ & $1$ & $0$ & $1$ &  &  &  &  &  &  &  &  &  &  &  &  &  &  &  & \\
$8$ & $2$ & $105$ & $2$ & $0$ & $3$ & $1$ & $3$ & $1$ & $2$ & $0$ & $1$ &  &  &  &  &  &  &  &  &  &  &  & \\
$8$ & $3$ & $462$ & $3$ & $1$ & $5$ & $4$ & $7$ & $4$ & $6$ & $3$ & $4$ & $2$ & $2$ & $0$ & $1$ &  &  &  &  &  &  &  & \\
$8$ & $4$ & $1530$ & $5$ & $2$ & $10$ & $7$ & $14$ & $10$ & $14$ & $9$ & $12$ & $7$ & $8$ & $4$ & $5$ & $2$ & $2$ & $0$ & $1$ &  &  &  & \\
$8$ & $5$ & $4191$ & $6$ & $5$ & $16$ & $14$ & $23$ & $20$ & $26$ & $21$ & $25$ & $19$ & $20$ & $14$ & $15$ & $9$ & $9$ & $5$ & $5$ & $2$ & $2$ & $0$ & $1$\\
$10$ & $1$ & $21$ & $0$ & $1$ & $0$ & $1$ & $0$ & $1$ &  &  &  &  &  &  &  &  &  &  &  &  &  &  & \\
$10$ & $2$ & $196$ & $2$ & $0$ & $4$ & $1$ & $4$ & $2$ & $3$ & $1$ & $2$ & $0$ & $1$ &  &  &  &  &  &  &  &  &  & \\
$12$ & $1$ & $28$ & $1$  & $0$ & $1$  & $0$  & $1$  & $0$  & $1$  &   &   &   &   &   &   &   &   &   &   &   &  &   & \\
$12$ & $2$ & $336$ & $3$  &  $0$ & $4$  & $2$  & $5$ & $2$ & $5$  & $2$  & $3$  & $1$  & $2$  & $0$  & $1$  &  &   &   &   &   &  &   & \\
$14$ & $1$ & $36$ & $0$  & $1$   & $0$  & $1$  & $0$  & $1$  & $0$  & $1$  &   &   &   &   &   &   &   &   &   &   &  &   & \\
\end{tabular}
\end{ruledtabular}
\caption{Number of multiplets of states at zero energy for the three-body Hamiltonian for  $\nu = 1$ for $\Delta\Phi$ added quantum fluxes. $\#$ is the total number of degenerate states.}
\label{tabledegeneracy1}
\end{table*}

\begin{table*}
\begin{ruledtabular}
\begin{tabular}{cccccccccccccccccccccc}
$N$ & $\Delta\Phi$ & $\#$ & $L=0$ & $1$ & $2$ & $3$ & $4$ & $5$ & $6$ & $7$ & $8$ & $9$ & $10$ & $11$ & $12$ & $13$ & $14$  & $15$ & $16$ & $17$ & $18$\\  \hline
$6$ & $1$ & $10$ & $0$ & $1$ & $0$ & $1$ &  &  &  &  &  &  &  &  &  &  &  &  &  &  &  \\
$6$ & $2$ & $50$ & $2$ & $0$ & $2$ & $1$ & $2$ & $0$ & $1$ &  &  &  &  &  &  &  &  &  &  &  &  \\
$6$ & $3$ & $165$ & $0$ & $2$ & $1$ & $4$ & $2$ & $3$ & $2$ & $2$ & $0$ & $1$ &  &  &  &  &  &  &  &  &  \\
$6$ & $4$ & $427$ & $3$ & $0$ & $4$ & $3$ & $6$ & $3$ & $6$ & $3$ & $4$ & $2$ & $2$ & $0$ & $1$ &  &  &  &  &  &  \\
$6$ & $5$ & $944$ & $0$ & $4$ & $2$ & $7$ & $5$ & $8$ & $7$ & $8$ & $5$ & $7$ & $4$ & $4$ & $2$ & $2$ & $0$ & $1$ &  &  &  \\
$6$ & $6$ & $1869$ & $4$ & $1$ & $7$ & $5$ & $11$ & $7$ & $13$ & $9$ & $12$ & $9$ & $10$ & $6$ & $8$ & $4$ & $4$ & $2$ & $2$ & $0$ & $1$ \\
$9*$ & $1$ & $20$ & $0$ & $1$ & $1$ & $0$ & $1$ &  &  &  &  &  &  &  &  &  &  &  &  &  &  \\
$9$ & $2$ & $175$ & $0$ & $3$ & $1$ & $5$ & $2$ & $3$ & $2$ & $2$ & $0$ & $1$ &  &  &  &  &  &  &  &  &  \\
$9*$ & $3$ & $870$ & $2$ & $6$ & $7$ & $8$ & $9$ & $9$ & $7$ & $8$ & $5$ & $4$ & $3$ & $2$ & $0$ & $1$ &  &  &  &  &  \\
$9$ & $4$ & $3122$ & $6$ & $5$ & $14$ & $14$ & $21$ & $17$ & $23$ & $18$ & $20$ & $16$ & $16$ & $10$ & $11$ & $6$ & $5$ & $3$ & $2$ & $0$ & $1$ \\
$12$ & $1$ & $35$ & $1$ & $0$ & $1$ & $1$ & $1$ & $0$ & $1$ &  &  &  &  &  &  &  &  &  &  &  &  \\
$12$ & $2$ & $490$ & $4$ & $1$ & $6$ & $4$ & $8$ & $4$ & $7$ & $3$ & $4$ & $2$ & $2$ & $0$ & $1$ &  &  &  &  &  &  \\
$15*$ & $1$ & $56$ &  $0$ & $1$ & $1$ & $1$ & $1$ & $1$ & $0$ & $1$\\
\end{tabular}
\end{ruledtabular}
\caption{Number of multiplets of states at zero energy for the four-body Hamiltonian for  $\nu = 3/2$ for $\Delta\Phi$ added quantum fluxes. $\#$ is the total number of degenerate states. $L$ values for rows with a star, have to be understood as $L-1/2$.}
\label{tabledegeneracy32}
\end{table*}

\begin{table*}
\begin{ruledtabular}
\begin{tabular}{cccccccccccccccccccccccc}
$N$ & $\Delta\Phi$ & $\#$ & $L=0$ & $1$ & $2$ & $3$ & $4$ & $5$ & $6$ & $7$ & $8$ & $9$ & $10$ & $11$ & $12$ & $13$ & $14$  & $15$ & $16$ & $17$ & $18$ & $19$ & $20$\\  \hline
$8$ & $1$ & $15$ & $1$ & $0$ & $1$ & $0$ & $1$ &  &  &  &  &  &  &  &  &  &  &  &  &  &  &  & \\
$8$ & $2$ & $105$ & $2$ & $0$ & $3$ & $1$ & $3$ & $1$ & $2$ & $0$ & $1$ &  &  &  &  &  &  &  &  &  &  &  & \\
$8$ & $3$ & $440$ & $2$ & $1$ & $4$ & $3$ & $6$ & $4$ & $6$ & $3$ & $4$ & $2$ & $2$ & $0$ & $1$ &  &  &  &  &  &  &  & \\
$8$ & $4$ & $1379$ & $4$ & $1$ & $7$ & $5$ & $11$ & $7$ & $12$ & $8$ & $11$ & $7$ & $8$ & $4$ & $5$ & $2$ & $2$ & $0$ & $1$ &  &  &  & \\
$8$ & $5$ & $3591$ & $4$ & $3$ & $10$ & $9$ & $16$ & $14$ & $19$ & $16$ & $20$ & $16$ & $18$ & $13$ & $14$ & $9$ & $9$ & $5$ & $5$ & $2$ & $2$ & $0$ & $1$ \\
$12$ & $1$ & $35$ & $1$ & $0$ & $1$ & $1$ & $1$ & $0$ & $1$ &  &  &  &  &  &  &  &  &  &  &  &  &  & \\
$12$ & $2$ & $490$ & $4$ & $1$ & $6$ & $4$ & $8$ & $4$ & $7$ & $3$ & $4$ & $2$ & $2$ & $0$ & $1$ &  &  &  &  &  &  &  & \\
$12$ & $3$ & $3311$ & $6$ & $6$ & $15$ & $16$ & $22$ & $19$ & $25$ & $20$ & $21$ & $17$ & $17$ & $11$ & $11$ & $6$ & $5$ & $3$ & $2$ & $0$ & $1$ &  & \\
$16$ & $1$ & $70$ & $1$ & $0$ & $2$ & $0$ & $2$ & $1$ & $1$ & $0$ & $1$ &  &  &  &  &  &  &  &  &  &  &  & \\
\end{tabular}
\end{ruledtabular}
\caption{Number of multiplets of states at zero energy for the five-body Hamiltonian for  $\nu = 2$ for $\Delta\Phi$ added quantum fluxes. $\#$ is the total number of degenerate states.}
\label{tabledegeneracy2}
\end{table*}

\begin{table*}
\begin{ruledtabular}
\begin{tabular}{cccccccccccccccccccc}
$N$ & $\Delta\Phi$ & $\#$ & $L=0$ & $1$ & $2$ & $3$ & $4$ & $5$ & $6$ & $7$ & $8$ & $9$ & $10$ & $11$ & $12$ & $13$ & $14$  & $15$\\  \hline
$10$ & $1$ & $21$ & $0$ & $1$ & $0$ & $1$ & $0$ & $1$ &  &  &  &  &  &  &  &  &  &  & \\
$10$ & $2$ & $196$ & $2$ & $0$ & $4$ & $1$ & $4$ & $2$ & $3$ & $1$ & $2$ & $0$ & $1$ &  &  &  &  &  & \\
$10$ & $3$ & $1001$ & $0$ & $4$ & $3$ & $7$ & $6$ & $9$ & $7$ & $9$ & $6$ & $7$ & $4$ & $4$ & $2$ & $2$ & $0$ & $1$ \\
$15^*$ & $1$ & $56$ & $0$ & $1$ & $1$ & $1$ & $1$ & $1$ & $0$ & $1$ &  &  &  &  & \\
$15$ & $2$ & $1176$ & $0$ & $7$ & $4$ & $12$ & $8$ & $12$ & $9$ & $11$ & $6$ & $8$ & $4$ & $4$ & $2$ & $2$ & $0$ & $1$ \\
\end{tabular}
\end{ruledtabular}
\caption{Number of multiplets of states at zero energy for the six-body Hamiltonian for  $\nu = 5/2$ for $\Delta\Phi$ added quantum fluxes. $\#$ is the total number of degenerate states. $L$ values for rows with a star, have to be understood as $L-1/2$.}
\label{tabledegeneracy52}
\end{table*}


\begin{table*}
\begin{ruledtabular}
\begin{tabular}{cccccccccc}
$N$ & ${\cal O}^{2,2}$ & ${\cal O}_0^{2,2}$ & ${\cal O}_1^{2,2}$ & ${\cal O}_2^{2,2}$ & ${\cal O}_3^{2,2}$ & ${\cal O}_4^{2,2}$ & ${\cal O}_5^{2,2}$ & ${\cal O}_6^{2,2}$ & ${\cal O}_7^{2,2}$  \\\hline
$4$ & $1.0$ & $1.0$ &  & $1.0$ \\
$6$ & $0.8579$ &  & $0.9994$ &  & $0.7972$ \\
$8$ & $0.8760$ & $0.9857$ &  & $0.8884$ &  & $0.8569$ \\
$10$ & $ 0.8661$ &  & $0.9111$ &  & $0.8334$ &  & $0.8287$ \\
$12$ & $0.6800$ & $0.9651$ &  & $0.8007$ &  & $0.8392$ &  & $0.5015$ \\
$14$ & $0.6825$ &  & $0.7685$ &  & $0.7059$ &  & $0.7417$ &  & $0.611$ \\
\end{tabular}
\end{ruledtabular}
\caption{Overlap of the lowest energy exact wavefunctions and the
Pfaffian two quasihole wavefunctions at $\nu=1$.}
\label{tableoverlap1}
\end{table*}

\begin{table*}
\begin{ruledtabular}
\begin{tabular}{cccccccccc}
$N$ & ${\cal O}^{3,3}$ & ${\cal O}_0^{3,3}$ & ${\cal O}_1^{3,3}$ & ${\cal O}_2^{3,3}$ & ${\cal O}_3^{3,3}$ & ${\cal O}_4^{3,3}$ & ${\cal O}_5^{3,3}$ & ${\cal O}_6^{3,3}$ & ${\cal O}_7^{3,3}$  \\\hline
$6$ & $1.0$ &  & $1.0$ &  & $1.0$ \\
$9^*$ & $0.7933$ &  & $0.9561$ & $0.7065$ &  & $0.7803$ \\
$12$ & $0.6289$ & $0.8221$ &  & $0.7317$ & $0.8579$ & $0.5735$ &  & $0.4897$ \\
$15^*$ & $0.4440$ &  & $0.7965$ & $0.5754$ & $0.5856$ & $0.4440$ & $0.7572$ &  & $0.0009$ \\
\end{tabular}
\end{ruledtabular}
\caption{Overlap of the lowest energy exact wavefunctions and the
$k=3$ three quasihole wavefunctions at $\nu=3/2$.
$L$ values for rows with a star, have to be understood as
$L-1/2$.} \label{tableoverlap32}
\end{table*}

\begin{table*}
\begin{ruledtabular}
\begin{tabular}{ccccccccccc}
$N$ & ${\cal O}^{4,4}$ & ${\cal O}_0^{4,4}$ & ${\cal O}_1^{4,4}$ & ${\cal O}_2^{4,4}$ & ${\cal O}_3^{4,4}$ & ${\cal O}_4^{4,4}$ & ${\cal O}_5^{4,4}$ & ${\cal O}_6^{4,4}$ & ${\cal O}_7^{4,4}$ & ${\cal O}_8^{4,4}$\\  \hline
$8$ & $1.0$ & $1.0$ & & $1.0$ & & $1.0$ \\
$12$ & $0.5282$ & $0.9938$ &   & $0.6555$ & $0.7601$   & $0.6714$ &  & $0.2194$\\
$16$ & $0.3697$ & $0.6691$ &  & $0.7779$ &  & $0.6376$ & $0.1112$ & $0.2387$ &  & $0.0957$
\end{tabular}
\end{ruledtabular}
\caption{Overlap of the lowest energy exact wavefunctions and the
$k=4$ four quasihole wavefunctions at $\nu=2$.}
\label{tableoverlap2}
\end{table*}

\begin{table*}
\begin{ruledtabular}
\begin{tabular}{cccccccccc}
$N$ & ${\cal O}^{5,5}$ & ${\cal O}_0^{5,5}$ & ${\cal O}_1^{5,5}$ & ${\cal O}_2^{5,5}$ & ${\cal O}_3^{5,5}$ & ${\cal O}_4^{5,5}$ & ${\cal O}_5^{5,5}$ & ${\cal O}_6^{5,5}$ & ${\cal O}_7^{5,5}$\\  \hline
$10$ & $1.0$ & & $1.0$ & & $1.0$ & & $1.0$ \\
$15^*$ & $0.5507$ & & $0.6998$  & $0.7666$ & $0.8554$   & $0.6399$ &  $0.8110$  & & $0.2827$\\
\end{tabular}
\end{ruledtabular}
\caption{Overlap of the lowest energy exact wavefunctions and the
$k=5$ five quasihole  wavefunctions at $\nu=5/2$. $L$
values for rows with a star, have to be understood as $L-1/2$.}
\label{tableoverlap52}
\end{table*}

\begin{table*}
\begin{ruledtabular}
\begin{tabular}{ccccccccccccccc}
$N$ & ${\cal O}^{2,4}$ & ${\cal O}_0^{2,4}$ & ${\cal O}_1^{2,4}$ & ${\cal O}_2^{2,4}$ & ${\cal O}_3^{2,4}$ & ${\cal O}_4^{2,4}$ & ${\cal O}_5^{2,4}$ & ${\cal O}_6^{2,4}$ & ${\cal O}_7^{2,4}$  & ${\cal O}_8^{2,4}$  & ${\cal O}_9^{2,4}$  & ${\cal O}_{10}^{2,4}$  & ${\cal O}_{11}^{2,4}$  & ${\cal O}_{12}^{2,4}$  \\\hline
$4$ & $0.9807$ & $1.0$ &  & $1.0$ &  & $0.9572$ \\
$6$ & $0.8541$ & $0.9998$ &  & $0.9741$ & $0.9764$ & $0.9369$ &  & $0.5590$ \\
$8$ & $0.8458$ & $0.9036$ & & $0.9407$ & $0.9197$ & $0.8155$ & $0.7370$ & $0.8620$ & & $0.8184$ \\
$10$ & $0.6692$ & $0.8991$ & & $0.8269$ & $0.7729$ & $0.8281$ & $0.7291$ & $0.8090$ & $0.4411$ & $0.5091$ & & $0.2898$ \\
$12$ & $0.5788$ & $0.7964$ &  & $0.7970$ & $0.8253$ & $0.7832$ & $0.7756$ & $0.5882$ & $0.5654$ & $0.5572$ & $0.4255$ & $0.4489$ &  & $0.0692$\\
\end{tabular}
\end{ruledtabular}
\caption{Overlap of the lowest energy exact wavefunctions and the
Pfaffian four quasihole wavefunctions at $\nu=1$.}
\label{tableoverlap12}
\end{table*}


\begin{thebibliography}{99}

\bibitem{Madison00}
K. W. Madison , F. Chevy, W. Wohlleben, and J. Dalibard,
Phys.Rev. Lett. \textbf{84}, 806 (2000);
F. Chevy, K. Madison, and J. Dalibard,
Phys. Rev. Lett. \textbf{85}, 2223 (2000)

\bibitem{Aboshaeer01}
J. R. Abo-Shaeer , C. Raman, J. M. Vogels, and W. Ketterle,
Science \textbf{292}, 476 (2001).

\bibitem{Wilkin98}
N. K. Wilkin, J. M. F. Gunn and R. A. Smith,
Phys. Rev. Lett. {\bf 80}, 2265 (1998)

\bibitem{Wilkin00}
N. K. Wilkin and J. M. F. Gunn,
Phys. Rev. Lett. {\bf 84}, 6 (2000)

\bibitem{Cooper01}
N. R. Cooper, N. K. Wilkin, and J. M. F. Gunn,
Phys. Rev. Lett. {\bf 87}, 120405 (2001).

\bibitem{Regnault03}
N. Regnault and Th. Jolicoeur,
Phys. Rev. Lett. {\bf 91}, 030402 (2003);
Phys. Rev. B{\bf 69}, 235309 (2004).

\bibitem{Laughlin83}
R. B. Laughlin,
Phys. Rev. Lett. {\bf 50}, 1395 (1983).

\bibitem{Chiachen05}
C. Chang, N. Regnault, Th. Jolicoeur and J. K. Jain,
Phys. Rev. A{\bf 72}, 013611 (2005).

\bibitem{Jain89}
J.K. Jain,
Phys. Rev. Lett. {\bf 63}, 199 (1989);
Physics Today {\bf 53}(4), 39 (2000);
Physica E{\bf 20}, 79 (2003).

\bibitem{Moore91}
G. Moore and N. Read,
Nucl. Phys. B{\bf 360}, 362 (1991).

\bibitem{Read99}
N. Read and E. H. Rezayi,
Phys. Rev. B{\bf 59}, 8084 (1999).

\bibitem{Rezayi05}
E. H. Rezayi, N. Read and N. R. Cooper,
Phys. Rev. Lett. \textbf{95}, 160404 (2005).

\bibitem{Ardonne02}
E. Ardonne,
J. Phys. A{\bf 35} 447 (2002).

\bibitem{Cappelli01}
A. Cappelli, L. S. Georgiev and I. T. Todorov,
Nucl. Phys. B{\bf 599}, 499 (2001).
\bibitem{Read96}
N. Read and E. H. Rezayi,
Phys. Rev. B{\bf 54}, 16864 (1996).

\bibitem{Zamolodchikov85}
A. B. Zamolodchikov and V. A. Fateev,
Sov. Phys. JETP {\bf 62}, 215 (1985).

\bibitem{Gurarie00}
V. Gurarie and E. H. Rezayi,
Phys. Rev. B{\bf 61}, 5473 (2000).

\bibitem{Gepner87}
D. Gepner and Z. Qiu,
Nucl. Phys. B{\bf 285}, 423 (1987).

\bibitem{Freedman03}
M. H. Freedman, A. Kitaev, M. J. Larsen, and Z. Wang,
Bull. AMS, {\bf 40}, 31 (2003).

\bibitem{Fano86}
G. Fano, F. Ortolani, and E. Colombo,
Phys. Rev. B\textbf{34}, 2670 (1986).

\bibitem{Wu76}
T. T. Wu and C. N. Yang,
Nucl. Phys. B{\bf 107}, 365 (1976);
Phys. Rev. D{\bf 16}, 1018 (1977).

\bibitem{Haldane83}
F. D. M. Haldane,
Phys. Rev. Lett. {\bf 51}, 605 (1983).

\end{thebibliography}
\end{document}